\documentclass{ws-procs975x65}            
\begin{document}
\title{Pair creation in the early universe}

\author{Stahl Cl\'ement}


\author{Strobel Eckhard and Xue She-Sheng}

\address{ICRANet, Piazzale della Repubblica, 10,  65122 Pescara, Italy\\
Dipartimento di Fisica, Universita di Roma "La Sapienza", Piazzale Aldo Moro 5, 00185 Rome, Italy \\
Universit\'e de Nice Sophia Antipolis, 28 Avenue de Valrose, 06103 Nice Cedex 2, France \\
E-mail: clement.stahl@icranet.org ;  xue@icra.it\\
}

\begin{abstract}
 In the very early universe, a generalized Schwinger effect can create pairs from both electrical and gravitational fields. The expectation value of fermionic current induced by these newly created pairs has been recently computed in de Sitter spacetime. I will discuss different limiting cases of this result and some of its possible physical interpretations.

\end{abstract}


\bodymatter

\section{Introduction}\label{aba:sec1}
The Schwinger effect: the creation of pair of particle/anti-particle under the influence of an external electric field, is a remarkable non-perturbative quantum effect. Since the original work of Heisenberg, his student Euler,\cite{Heisenberg1936} Sauter\cite{Sauter1931} and Schwinger,\cite{Schwinger1951} this effect has never been measured due to the need for huge electric field ($E_\text{critical}=10^{16} V/cm$). One fruitful playground to investigate Schwinger pair creation in more general cases is the early universe. Strong electric and gravitational fields could be naturally present and would both allow virtual pairs to tunnel from the Dirac sea.

To account correctly for gravitational effects in early universe, one needs to consider Quantum Field Theory in curved spacetime. One of the simplest curved spacetime is de Sitter spacetime which is believed to be of great importance both for inflationary scenarii and late time acceleration of the expansion of the universe. In the recent literature, these generalizations of the Schwinger effect in curved spacetime were investigated and permit to be conclusive on many theoretical problems: bubble nucleation,\cite{Froeb2014} testing the conjecture ER=EPR,\cite{Fischler2014} constraining magnetogenesis,\cite{Kobayashi2014} connecting Hawking and Unruh effect to Schwinger effect.\cite{Sang1,Sang2}

Whereas most of these works focus on scalar Quantum Electrodynamics (QED), spinor QED results were derived in the last months.\cite{Stahl2015A,Stahl2015B} This step makes possible an explicit comparison between boson and fermion creation rate for constant electric field and the interesting fact that boson and fermion are not created with the same rate under the influence of an electric field was discovered.

I discussed in my presentation two calculations:\cite{Stahl2015B} a semiclassical estimate for the number of fermion created in 2D de Sitter spacetime ($dS_2$) and different limiting cases of the induced current. The derivation of this induced current has been discussed in details in the presentation of my collaborator Eckhard Strobel in the same parallel session and will be also reviewed for the sake of completeness. I will finish this proceeding by discussing two physical possible consequences: the debated issue of the new effect found originally in Ref.~\citenum{Froeb2014}: infrared conductivity and an attempt to solve the matter/anti-matter asymmetry problem with the help of the multiverse proposal.

\section{Dirac field in $dS_2$: generalities.}
We start from the action for QED in curved spacetime coupled to a spinor field \(\psi(x)\)
$$
\label{eq:action}
S=\int \text{d}^2 x  \sqrt{-g(x)} \left[  -\frac{1}{\kappa}R(x) +\frac{i}{2} \left[\overline{\psi}(x) \underline{\gamma}^{\mu} \nabla_{\mu} \psi(x) -\left(\nabla_{\mu} \overline{\psi}(x)\right)\underline{\gamma}^{\mu}\psi(x)\right] -m \overline{\psi}(x) \psi(x)  -\frac{1}{4} F_{\mu \nu}(x)F^{\mu \nu}(x) \right].
$$
The electromagnetic side and gravitational side of the action are considered as non-dynamical \emph{ie.} backreaction effects of the fermionic field to the gravity and electromagnetic sector are neglected. The electric field is taken to be constant and the metric is chosen as de Sitter. Varying the action with respect to the fermionic field gives the Dirac equation which can be solved with the help of the Whittacker functions. The positive and negative frequency solutions can then be constructed and related via charge conjugation.
\section{Bogoliubov calculation}
With explicit solutions, the standard Bogoliubov calculation can be performed.\cite{Froeb2014,Kobayashi2014} It relies on relating solutions in the asymptotic past to solutions in the asymptotic future with the help of the Bogoliubov transformation:
$$
 \psi_\text{in}^+(\eta)=\alpha_k \,\psi_\text{out}^+(\eta)+\beta_k\, \psi_\text{out}^-(\eta) \label{eq:inout}
$$
where the superscript + and - are positive and negative frequency solutions respectively. The number of created pairs per mode can then be derived with the help of a connection formula between the Whittacker functions $\psi_\text{out}^+(\eta)$ and $\psi_\text{in}^+(\eta)$, it is found to be:\cite{Stahl2015B}
$$
n_k=|\beta_k|^2=e^{-\pi (\mu- r \lambda)}\frac{\sinh(\pi(\mu+ r \lambda))}{\sinh(2\pi \mu)},\label{eq:nk}
$$
where $r \equiv \frac{k}{|k|}$ is the sign of the momentum, $\lambda \equiv \frac{eE}{H^2}$ dimensionless electrical force and $\mu \equiv \sqrt{\lambda^2 + \frac{m^2}{H^2}}$.
To derive this result, one needs the notion of particle in the asymptotic future. To have it, one needs that the change of the background to be small in the asymptotic future which gives after calculations $\mu \gg 1$.

 Furthermore, it is important to note this number of created pair agrees with the number of created pair derived by the semiclassical scattering method which was also presented in our paper.\cite{Stahl2015B} The semiclassical scattering method is a technique to calculate the number of pairs for very general electrical field configurations. In Eckhard Strobel's PhD thesis,\cite{Strobel} it was well explained that it relies ultimately on approximating integrals by the value of their residues at the turning points. It is usually referred as to WKB but it is more precise\cite{Strobel,Blinne:2015zpa} to called it the \emph{semiclassical scattering method}.  The semiclassical scattering method has been successfully applied in flat spacetime problems for a two-components electrical field\cite{Strobel:2014tha} as well as in 4D and 3D \cite{Barbar} de Sitter spacetime\cite{Stahl2015A} for any one-component electrical field. 

\section{The induced current: generalities}
As discussed in the previous section, the number of created pairs does need the notion of particle to be well defined. In curved spacetime the fact that there is no more $\frac{\partial}{\partial_0 }$ Killing vector implies that the vacuum state of the Fock space is not unique anymore. Hence the very notion of particle is not uniquely defined in the general case. To overcome these difficulties, we considered a more general quantity: the induced current which coincides with the number of particle in the semiclassical limit ($\mu \gg 1$). It is defined as:
$$
 J^x=\frac{e}{2}\left[\overline{\psi}(x),{\gamma}^x\psi(x)\right]
$$
By using an explicit set of solutions \emph{eg.} $\psi_\text{in}^-(\eta)$ and $\psi_\text{in}^+(\eta)$ and after an involved integration procedure in the complex plane,\cite{Stahl2015B} it is possible to find the following result:
$$
 J^x=\frac{eH}{\pi}\left(\mu\frac{\sinh(2\pi\lambda)}{\sin(2\pi\mu)}+\lambda\right).
 \label{eq:nonnormcurr}
$$
However, this current needs to be renormalized. In $dS_4$, for bosonic particles, an adiabatic subtraction scheme was used. In $dS_2$, for bosonic particles a Pauli-Villars renormalization scheme was used.\cite{Froeb2014} It can be shown that in $dS_2$ for boson, the Pauli-Villars and the adiabatic subtraction methods agree\footnote{Private communications between Eckhard Strobel and Cl\'ement Stahl}.

The adiabatic subtraction relies on a reparametrization of the Dirac equation. For bosonic particle, it is simply done with a WKB ansatz with a free function for the effective frequency. However for fermionic particle, this WKB ansatz is not the right one anymore, we had to generalize two works. The correct ansatz was already found in flat spacetime with a constant electrical field\cite{Kluger1992} and in curved spacetime without electrical field.\cite{Landete2014} The new ansatz that we introduced is:
$$
 \psi^+(\eta)=N\sqrt{\frac{1}{2\Omega(\eta)}}\exp\left(\int^\eta\left[-i\Omega(t)+\frac{p(t)}{2\Omega(t)}\left(\frac{p'(t)}{p(t)}-\frac{a'(t)}{a(t)}\right)\right]dt\right),
$$
where the kinetical momentum is $p(\eta) \equiv k + eA(\eta)$ and the scale factor is $a(\eta)=\frac{-1}{H\eta}$ in de Sitter spacetime.

Following the adiabatic subtraction procedure,\cite{red book} we find\cite{Stahl2015B} up to forth adiabatic order:
$$
  J^x=\frac{eH}{\pi}\lambda+O\left(T^{-4}\right)
$$
It cancels exactly the second term of (\ref{eq:nonnormcurr}). Those counter-terms and their associated divergence depends on the number of spatial dimension.

To better understand those divergences, we are now studying 3D bosonic pair creation,\cite{Barbar} the ultimate goal being knowing the relation between the different renormalization schemes in curved spacetime.
\section{Limiting cases for the induced current}
The derivation of the fermionic current was the object of Eckhard Strobel's presentation. For the sake of completeness, it has been depicted with few details in the previous section, the final result is:\cite{Stahl2015B}
$$
J^x=\frac{eH}{\pi}\mu\frac{\sinh\left(2\pi\lambda\right)}{\sinh(2\pi\mu)}.
$$
 \begin{figure}
 \centering
 \includegraphics[width=1\textwidth]{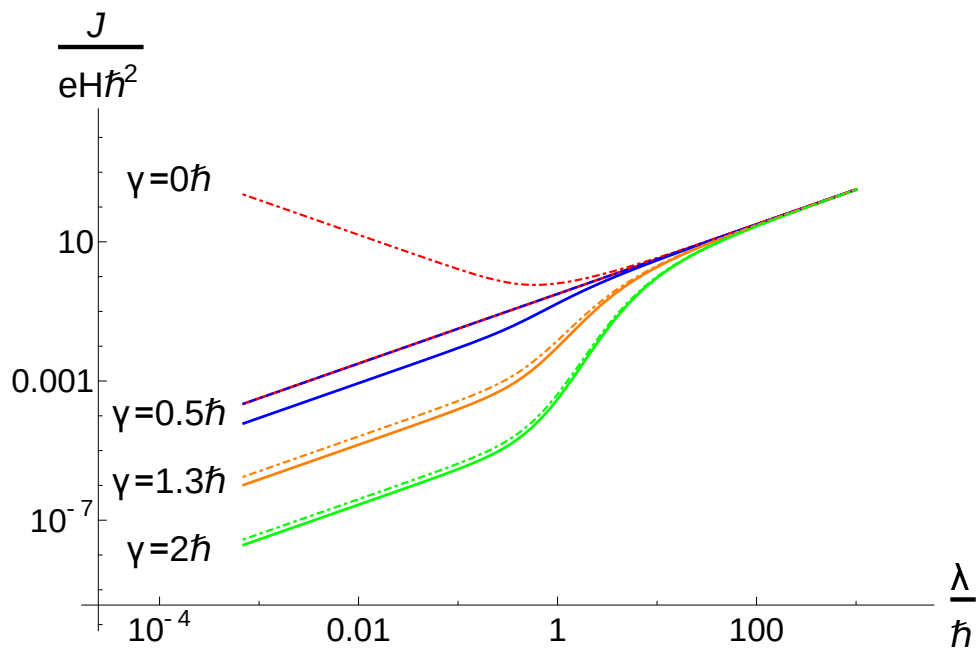} 
 \caption{Regularized current  for fermions (solid lines) and bosons (dotted lines) respectively as a function of the electric field strength \(\lambda\) for different values of the mass over the Hubble constant: \(\gamma\equiv \frac{m}{H}\). There are differences between bosons and fermions for small electric fields whereas they have the same asymptotic limit for negligible mass. Only the bosonic case shows the ``infrared hyper-conductivity", \emph{ie.}~a large current for small electric field and mass. The linear response is for \(\gamma=0\) in the fermionic case and for \(\gamma=0.5\hbar\) in the bosonic case.}
 \label{fig:current}
\end{figure}

The followings properties can be shown analytically and agree with the numerical results:
\begin{itemize}
\item It agrees with the semiclassical estimates of the previous section in the semiclassical limit ($\mu \gg 1$: which implies $\lambda \gg 1$ because the relation between current and number of pairs holds for relativistic carrier \emph{ie.} $m \ll H$) as well as the estimates based on the scattering method.\cite{Stahl2015B}
\item One recovers the usual Schwinger effect of flat spacetime in the limit $H \rightarrow 0$. For the massless case, the fermionic field is conformally invariant and one also finds the usual Schwinger rate.
\item It is independent of the mass for strong electric field and has the conformal linear behavior (see also Fig.\ref{fig:current}).
\item It also has a linear behavior for weak field and there is no presence of IR-hyperconductivity in contrary to the bosonic case.\cite{Froeb2014} The standard picture of heavier particle harder to move to create a current is the physical interpretation of this result.
\item It is equivalent to the bosonic case with the change $\mu^2 \rightarrow \mu^2 - 1/4$. We will see in the next section that it is possible to perform this change by introducing a conformal coupling in the action.
\end{itemize}
Moving forward, I will present in the next section two ideas to use the pairs created by the electrical and gravitational fields during inflation.
\section{Discussion and speculations}
\subsection{Regarding infrared hyperconductivity}
In the work presented in the previous sections, no infrared hyperconductivity has been reported which lets us think that the fermionic case is less peculiar than the bosonic case. Indeed, for fermions the linear response which is a flat spacetime behavior is recovered for $m=0$ signaling that a massless fermion is conformally invariant whereas for bosons, the linear response is found for $m=0.5H$ and for $0 \leq m<0.5H$, the phenomenon of infrared-hyperconductivity occurs.  

Our new paper\cite{Barbar} aims at clarifying those issues. The first point is that by introducing a conformal coupling ie. a term in the action of the type $\xi R(x) \varphi(x)^* \varphi(x)$ with $\varphi(x)$ a complex spin-0 field, the  only change is that the mass of the field becomes $m^2 \rightarrow m^2 + 2 \xi H^2$. Non minimally coupled theories are well motivated extensions of General Relativity and are also motivated from string theory.\cite{Kachru:2003sx} Looking at $\xi$ as a free parameter of this new model, it is easy to see that setting $\xi=1/8$, the bosonic and the fermionic case would agree and no infrared hyperconductivity would occur for  $m^2 \geq 0$. But it would for $m^2<0$ revealing that in these beyond general relativity, infrared hyperconductivity would be a tachyonic excitation of the bosonic field. Its relation to general relativity through conformality remains to be investigated. The other way round, for $\xi>1/8$, there would be infrared hyperconductivity for $\frac{m^2}{H^2} < 1/4-2\xi$, which shows that infrared hyperconductivity is a common feature of beyond General Relativity theories.

Whereas for $m=0$, it was impossible to draw any conclusion in 4D de Sitter spacetime because the renormalization procedure was not appropriate, in 2D\cite{Froeb2014} and in 3D,\cite{Barbar} the infrared hyperconductivity was found to be maximal. In both cases, the current was unbounded from above and increased for decreasing electrical field.
Investigating infrared hyperconductivity in 1+1 D de Sitter spacetime, preliminary results seems to show that a regime of infrared hyperconductivity could be an interesting way of amplifying a spin-1 field from vacuum fluctuation\cite{SSS} which is the preferred primordial magnetogenesis scenario. Those models, if confirmed in 4D, would be different from the already existing ones in the sense that they do not require to break the conformal invariance of the Maxwell equation neither in the kinetic term nor in the potential term but they only add a current which couples to the spin-1 field.
\subsection{Regarding baryogenesis}
We are planning in a future paper\cite{future} to apply all these models treating Schwinger effect to baryogenesis scenario. I will develop in advance a few thoughts along this idea. The very idea is to separate the pairs particle/anti-particle one from each other with the help of strong gravitational and electrical fields present during inflation. However one needs to define properly a charge operator $\hat{C}$. The quantum fields theories manipulated in the previous section can be checked to be $\hat{C}\hat{P}\hat{T}$ invariant. The charge operator has been identified as acting in the following way for a fermion:\cite{Stahl2015B} $\hat{C}\psi(x) \equiv i \sigma_2 \psi^*(x)$. Looking explicitly at the solutions in this model, one would recover once again the Feynman physical picture of anti-particle traveling backwards in time. Those solutions for particles and anti-particles behave differently and it is indeed possible to observe a charge violation.

Hence the next steps would be to consider a quasi de Sitter spacetime to allow a reheating phase and to move to 3 spatial dimension. These technical and mathematical works being done, the physical picture would be that virtual pairs could tunnel from the Dirac sea and become real and stretched to large scales because by the accelerated, non causal, expansion of the universe. Then the electrical field would sort them in a way that matter would aggregate in a given spatial region whereas an anti-matter would aggregate in another region.

The theory of inflationary multiverses is based on melting inflationary cosmology, anthropic considerations, and particle physics. A very recent paper\cite{Linde} describes the growth of this theory from its infancy. The very idea of inflation was to render our part of the Universe homogeneous by stretching any pre-existing inhomogeneities to scales inaccessible to us. If our universe consists in several parts, each of these parts after inflation will become locally homogeneous and inhabitants of a given part will not be able to communicate with any other part and will conclude wrongly that the universe is homogeneous everywhere. These very large parts are sometimes called \emph{mini-universes}\cite{Linde} or \emph{pocket universes}.\cite{Carr} Finally the whole system consisting of many pocket universes is called \emph{multiverse} or \emph{inflationary multiverse}.

This picture of eternal inflation mays sound speculative but its ingredients are well rooted into theoretical physics' tools. It relies on three pillars: first, the non-uniqueness of vacuum state. It is possible in the standard model of particle physics, common in beyond standard model physics and inevitable in string theory. Second eternal inflation is a theory based on a quantum field theory framework which is so far the main way to do theoretical physics. Third, eternal inflation needs an accelerated phase of expansion. A late time expansion  (dark energy) was already observed and the early time inflation is inferred from cosmic microwave background observation for instance. If one believes these three pillars, eternal inflation is a direct natural consequence as described in the previous paragraph.

Particles created by Schwinger mechanism in de Sitter spacetime were already used in Ref.~\citenum{Froeb2014} to model bubble nucleation in de Sitter spacetime in the context of inflationary multiverses. Within those models, it is possible that the transition via an Higgs mechanism from an inflationary universe to the standard reheating phase does not occur simultaneously everywhere. One could imagine that the phase transition of the two zones created by Schwinger effect described in the previous paragraphs would not occur at the same time and hence form two causally disconnected patches: two pocket universes. One only filled with matter, the other only filled with anti-matter. This could be a proposal to explain the matter/anti-matter asymmetry, see also the discussion of Ref.~\citenum{Linde} about baryogenesis proposals. To my best knowledge, this idea of using particle created by gravitational and electrical Schwinger effect to solve the baryogenesis problem is a new one. See also Ref.~\citenum{Goolsby-Cole:2015chd} for a topical discussion of the charge of the universe during inflation and connection to baryogenesis.

 For our scenario to occur, one needs to assume a given number of premises. First that the decay of the particles created during reheating preserves the baryon number so that the already present asymmetry stays. Second that the expansion of the universe, which can still be accelerated in many models of preheating or for warm inflation would not dilute the already present asymmetry. Third that all the other ways of creating asymmetry are negligible regarding the main one. Fourth that there is a range of parameter in this model which predicts the value of the parameter $\eta\equiv \frac{n_b}{n_{\gamma}}=10^{-10}$. If all these requirements appear to be satisfied then, the three Sakharov conditions would be fulfilled and this scenario would be a valuable attempt to solve the baryogenesis problem. 

\section{Conclusion and perspectives}
Many generalizations of Schwinger effect in curved spacetime have been under scrutiny in the past years. Two calculations on this topic were sketched in my presentation and in this proceeding. The main guideline was given. One was a Bogoliubov calculation to derive the number density for the fermions created by a constant electric field in de Sitter spacetime. The other was the full derivation of the induced current. The result agrees with all the existing literature and gives an interesting difference between bosons and fermions. This difference seems to come from curvature effects, because there is no occurrence of a difference between boson and fermion in flat spacetime for constant electric field. In the discussion, I showed that if one introduces a conformal coupling, it is possible to link the boson and the fermion results and to reinterpret the regime of infrared hyperconductivity as a tachyonic excitation of the field. We plan to come back to the relation between infrared hyperconductivity, conformality and tachyonicity. I furthermore presented some ideas about infrared hyperconductivity, baryogenesis and the multiverse.

Generalizing this work to 4D can be an extension but we did not find yet solutions to the 4D Dirac equation. Considering backreaction effects (\emph{ie.} assuming  the newly created pairs can change the background electric and gravitational field vie Maxwell and Einstein equation) could also lead to new cosmological constrains as well as possible preferred metric or electric field to create pairs. I am now investigating those issues. Regarding backreaction to the gravitational sector of the theory, the de Sitter spacetime would not have a constant Hubble factor and the evolution of $H(t)$ could give clues about the possible relation between matter and dark energy,\cite{Xue:2014kna} more specifically, one could speculate that it would be the pair created by Schwinger effect that would convert the potential energy of the inflaton into kinetical energy and unleash the reheating.

\section*{Acknowledgements}
I thank Eckhard Strobel for a fruitful collaboration and Bernard Fricker for hosting me during a part of this work. I am supported by the Erasmus Mundus Joint Doctorate Program by Grant Number 2013-1471 from the EACEA of the European Commission.

\end{document}